\documentclass[pra,aps,twocolumn,amsmath,amssymb,superscriptaddress]{revtex4-1}
\usepackage{epsfig,amsmath}
\usepackage{subfigure}
\usepackage{graphicx}% Include figure files
\usepackage{dcolumn}% Align table columns on decimal point
\usepackage{stmaryrd}
\usepackage{mathrsfs}
\usepackage{pifont}
\usepackage{amsthm}
\usepackage{amssymb}
\usepackage{bm}
\usepackage{latexsym}
\usepackage[colorlinks=true,linkcolor=blue,citecolor=blue]{hyperref}
\usepackage{color}
\usepackage{epstopdf}
\usepackage[ruled]{algorithm2e}

\begin{document}

\title{Geometric quantum gates via dark paths in Rydberg atoms}

\author{Zhu-yao Jin}
\affiliation{School of Physics, Zhejiang University, Hangzhou 310027, Zhejiang, China}

\author{Jun Jing}
\email{Email address: jingjun@zju.edu.cn}
\affiliation{School of Physics, Zhejiang University, Hangzhou 310027, Zhejiang, China}

\date{\today}

\begin{abstract}
Nonadiabatic holonomic quantum gates are high-speed and robust. Nevertheless, they were found to be more fragile than the adiabatic gates when systematic errors become dominant. Inspired by the dark-path scheme that was used to partially relieve the systematic error in the absence of external noise, we construct a universal set of nonadiabatic holonomic $N$-qubit gates using the Rydberg-Rydberg interaction between atoms under off-resonant driving. Based on an effective four-level configuration in the Rydberg-atom system, the modified nonadiabatic holonomic geometric gates present a clear resilience to both systematic error in the whole parametric range and external noise. In our scheme, the conventional ultrastrong interaction between control atoms and the target atom for the nonadiabatic holonomic quantum computation is compensated by the detuning of the driving fields on the target atom. That idea yields a deeper understanding about the holonomic transformation. Moreover, our scheme is compact and scale-free with respect to $N$. It is interesting to find that the three-qubit gate is less susceptible to errors than the double-qubit one.
\end{abstract}

\maketitle

\section{Introduction}

Controlled gates are basic elements in the quantum circuit and are widely applicable in quantum error correction~\cite{Shor1995Scheme,Steane1996Error} and quantum algorithms~\cite{Shor1997Polynominal,Grover1997Quantum,
Grover1998Quantum,Vandersypen2001Experimental,Joshi2006Three,Yang2007Implementation}. In comparison to the quantum circuit composed of many one-qubit gates~\cite{Barenco1995Elementary} and two-qubit gates~\cite{Urban2009Observation,Barenco1995Elementary,Moller2008Quantum,Saffman2009Efficient}, the utilization of $N$-qubit controlled gates~\cite{Barenco1995Elementary,Moller2008Quantum,Saffman2009Efficient,Zheng2013Implementation} can reduce the number of quantum gates and then lead to more convenient and faster quantum information processing.

Many setups were used to realize controlled gates, such as trapped ions~\cite{Monroe1995Demonstration,Schmidt2003Realization,Monz2009Realization}, superconducting circuits~\cite{Yamamoto2003Demonstration,Plantenberg2007Demonstration,Fedorov2012Implementation}, linear optics~\cite{Brien2003Demonstration,Pittman2003Experimental,Miifmmode2013Efficient}, and neutral atoms~\cite{Graham2019Rydberg,Isenhower2010Demonstration,Levine2018High,Saffman2010Quantum,Jaksch2000Fast,
Petrosyan2017High,Gambetta2020Engineering,Gambetta2020Longrange,Young2021Asymmetric}. Among them, the neutral atoms have remarkable advantages with a long lifetime and a strong Rydberg-Rydberg interaction (RRI), i.e., the dipole-dipole or van der Waals interaction between the Rydberg states. By virtue of RRI, the neutral atoms demonstrate two opposite phenomena in dynamics which are useful in quantum control: (i) the Rydberg blockade~\cite{Lukin2001Dipole,Muller2009Mesoscopic,Gaetan2009Observation}, in which once an atom is excited to the Rydberg state, then the other atoms within the blocking radius will be inhibited from being excited, and (ii) the Rydberg antiblockade~\cite{Ates2007Antiblockade,Li2013Nonadiabatic,Pohl2009Breaking,Su2016Onestep}, in which the RRI is compensated by the detuning of laser fields and then more than one atom can be excited. Though two- and multiqubit gates based on RRI have been realized in experiments~\cite{Graham2019Rydberg,Wilk2010Entanglement,Zeng2017entangling,Maller2015Rydberg,Picken2019Entanglement,
Madjarov2009High}, they are still subject to the inevitable environmental noises and systematic errors.

Noise-resilient quantum gates~\cite{Solinas2004Robustness,Zhu2005Geometric,Solinas2012On,Johansson2012Robustness} can be generated by using the geometric phase~\cite{Wilczek1984Appearance,Berry1984Quantal,Aharonov1987phase,Pachos1999NonAbelian} that relies on the global rather than the local properties of the evolution path. The adiabatic geometric gates~\cite{Duan2001Geometric,Wu2005Holonomic,Huang2019Experimental} based on either Abelian~\cite{Berry1984Quantal} or non-Abelian~\cite{Wilczek1984Appearance} phases, however, require an adiabatic (slow) evolution. Considerable errors will otherwise be accumulated and then give rise to undesired transition and decoherence. Fast evolution could be realized by nonadiabatic holonomic quantum computation (NHQC)~\cite{Sjoqvist2012Nonadiabatic}. It was further improved by using the decoherence-free subspace~\cite{Xu2012Nonadiabatic,Sun2022Onestep}, the single-shot-shaped pulses~\cite{Xu2015Nonadiabatic}, a single-loop path~\cite{Herterich2016Singleloop}, and the dynamical decoupling technique~\cite{Genov2017Arbitrarily,Zhao2021Dynamical}. Among the NHQC variants that have been implemented in experiments~\cite{Abdumalikov2001Experimental,Xu2018Singleloop,Feng2013Experimental,Zu2014Experimental,Sekiguchi2017Optical}, the geometric gates have been found to be sensitive to the systematic errors~\cite{Zheng2016Comparison,Jing2017NonAbelian} from the imperfect state preparation and operation~\cite{Galindo2002Information,Nigg2014quantum}. To overcome this weakness, nonadiabatic holonomic quantum computation with two dark paths (NHQCTD)~\cite{Ming2022Experimental,Andr2022Dark} has been proposed with the dressed-state technique~\cite{Baksic2016Speeding}. In the closed-system scenario for the single-qubit~\cite{Ming2022Experimental} and single-qutrit gates~\cite{Andr2022Dark}, the dark-path scheme prevails over the standard NHQC method in resistance to the global error in Rabi frequency. In the open-system scenario~\cite{Ming2022Experimental,Andr2022Dark}, a tiny global error in Rabi frequency would however result in the NHQCTD scheme being worse than NHQC scheme in gate performance.

In this paper we construct nonadiabatic holonomic gates with two dark paths (NHGTD) in an $N$-partite system by the strong and stable Rydberg-like interaction between atoms. In our scheme, the typical resonant driving fields on the target atom~\cite{Yin2020Onestep,Sun2021Onestep} in NHQC are replaced with the off-resonant driving fields, which compensate for the ultrastrong interactions between the control and target atoms. Distinct from the dark-path schemes in the trapped-ion systems~\cite{Ming2022Experimental,Andr2022Dark}, our scheme shows global advantages in resisting systematic errors over the NHQC scheme since it is free of the ancillary level as well as the relevant unwanted leakage. In particular, we focus on an effective four-level configuration in a multiple-Rydberg-atom system through tuning the off-resonant driving fields on the target atom and one of the control atoms. The effective time-evolution operator $U(\tau)$ can be obtained by the dark-path method rather than the time-ordered integral in NHQC. In our scheme, one dark path is provided by the dark state of the system Hamiltonian with zero eigenvalue, which is decoupled from the dynamical process. Another one is obtained by the time-dependent Schr\"odinger equation with a vanishing expectation value about the effective Hamiltonian. The two dark-path states and the other two states out of the computational space constitute a completed space for the system evolution. Moreover, it is interesting to find that the $N$-qubit gates in our scheme can be realized with one instead of approximately $N$ steps~\cite{Xing2021Realization}.

The rest of the paper is structured as follows. In Sec.~\ref{modelandHam_two} we derive the effective Hamiltonian in a double-Rydberg-atom system for our NHGTD scheme, which generates a universal set of nonadiabatic holonomic two-qubit controlled gates. We verify the robustness of both controlled-NOT (CNOT) and controlled-Z (CZ) gates against the systematic errors and the external decoherence. In Sec.~\ref{modelandHam_three} we provide the effective Hamiltonian for the three-qubit controlled gates and demonstrate the robustness of the controlled-controlled-NOT (CCNOT) gate. In Sec.~\ref{Nqubit} our scheme is generalized to $N$-qubit controlled gates, which is exemplified with $N=4$. We summarize the paper in Sec.~\ref{Conclusion}.

\section{Nonadiabatic holonomic two-qubit controlled gates}\label{modelandHam_two}

\subsection{Gate construction with dark paths}

\begin{figure}[htbp]
\centering
\includegraphics[width=0.85\linewidth]{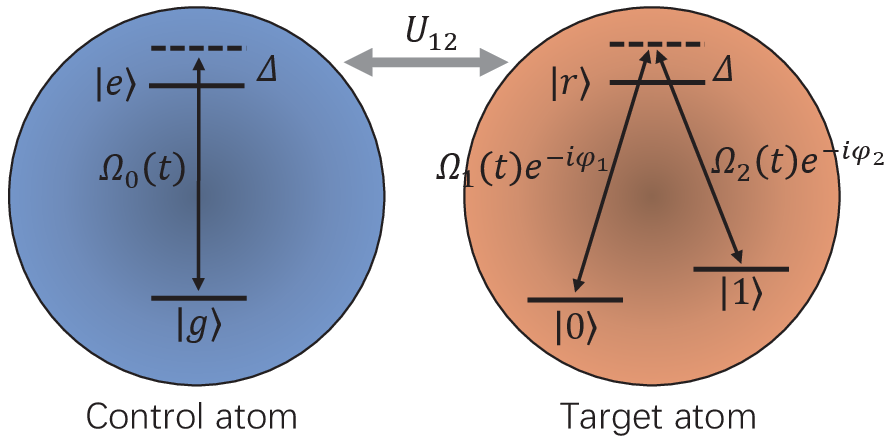}
\caption{Sketch of two coupled Rydberg atoms under off-resonant driving. The control atom consists of the ground state $|g\rangle$ and the Rydberg state $|e\rangle$. The target atom consists of two ground states $|0\rangle$ and $|1\rangle$ and one Rydberg state $|r\rangle$. Here $U_{12}$ is the Rydberg-like interaction between atoms.}\label{model}
\end{figure}

Consider two coupled Rydberg atoms under driving by three laser fields as shown in Fig.~\ref{model}. The control atom consists of the ground state $|g\rangle$ and the Rydberg state $|e\rangle$. The target atom has two stable low-energy states $|0\rangle$ and $|1\rangle$ and one Rydberg state $|r\rangle$. The control atom is driven by an off-resonant laser with detuning $\Delta$ and time-dependent Rabi frequency $\Omega_0(t)$. In the target atom, the transitions $|0\rangle\leftrightarrow|r\rangle$ and $|1\rangle\leftrightarrow|r\rangle$ are driven by the laser fields $\Omega_1(t)$ and $\Omega_2(t)$, respectively, with the same detuning $\Delta$; $\varphi_1$ and $\varphi_2$ are two time-independent initial phases. The two atoms are actually of the same type. The transitions associated with the state $|1\rangle$ of the control atom do not contribute to our scheme. The states $|0\rangle$ and $|r\rangle$ for the control atom are labeled with $|g\rangle$ and $|e\rangle$, respectively; hence the control atom can be distinguished from the target atom.

In the interaction picture with respect to the free Hamiltonian of the two atoms, the full Hamiltonian can be written as ($\hbar\equiv1$)
\begin{equation}\label{H}
H(t)=H_c(t)\otimes\mathcal{I}_t+\mathcal{I}_c\otimes H_t(t)+U_{12}|er\rangle\langle er|,
\end{equation}
with
\begin{equation}\label{Hsingle}
\begin{aligned}
H_c(t)&=\Omega_0(t)e^{-i\Delta t}|e\rangle_{c}\langle g|+{\rm H.c.},  \\
H_t(t)&=\Omega_1(t)e^{-i(\Delta t+\varphi_1)}|r\rangle_t\langle 0|\\
&+\Omega_2(t)e^{-i(\Delta t+\varphi_2)}|r\rangle_t\langle 1|+{\rm H.c.},
\end{aligned}
\end{equation}
where $\mathcal{I}_c$ and $\mathcal{I}_t$ represent the identity operators for the control and target atoms, respectively. The Rydberg-like interaction $U_{12}|er\rangle\langle er|$ in Eq.~(\ref{H}) plays a key role in our scheme, which could result from an energy shift of the operator $|er\rangle\langle er|$~\cite{Walker2008Consequences}. The coupling strength $U_{12}$ can be obtained by either the perturbation theory~\cite{Walker2008Consequences} for a comparatively small quantum principle number $n$ or a more accurate energy-level analysis~\cite{Mack2011measurement,Cano2014Multiatom} for a general $n$. It could be in the form of the van der Waals interaction (scaling with $d_{12}^{-6}$) or the dipole-dipole interaction (scaling with $d_{12}^{-3}$), where $d_{12}$ is the atomic distance. Typically, when $d_{12}$ is in the range of $0.5-5\mu$m~\cite{Walker2008Consequences}, the interaction is in the dipole-dipole interaction form, and a closer separation indicates a stronger interaction as well as a larger interaction fluctuation induced by the position fluctuation~\cite{Robicheaux2021Photon}. The magnitude of the position fluctuation $\delta d_{12}$ between atoms is proportional to the duration of the laser applied on the atom and inversely proportional to the wavelength of the laser and the mass of the atom~\cite{Robicheaux2021Photon}. To achieve a stable and strong interaction, the atoms could be sufficiently separated and excited to a high-lying Rydberg state $|r\rangle=|103S,J=1/2,m_J=1/2\rangle$~\cite{Cano2014Multiatom,Sun2021Onestep,Sun2022Onestep}. When $d_{12}>5\mu$m, the dominant interaction is in the van der Waals form $U_{12}=C_6/d_{12}^6$, where the coefficient could be as large as $C_6/2\pi=1.043\times10^5$ GHz $\mu m^6$. When $d\approx9 \mu$m, we could have a van der Waals interaction strength $U_{12}/2\pi\approx200$ MHz and the interaction fluctuation is about $\delta U_{12}/2\pi\sim0.1$ MHz for $\delta d_{12}\sim10^{-3} \mu$m~\cite{Robicheaux2021Photon}. When $d_{12}=5.54\mu m$, the van der Waals interaction could be as strong as $U_{12}/2\pi\approx3600$ MHz~\cite{Sun2021Onestep}. We therefore set $U_{12}/2\pi=500$ MHz in the following theoretical analysis. In an optical lattice~\cite{Gambetta2020Engineering} or ion crystals~\cite{Zhang2020Submicrosecond,Gambetta2020Longrange}, the Rydberg atoms can be strongly confined within a harmonic potential via phonon mode~\cite{Zhang2020Submicrosecond,Gambetta2020Engineering,Gambetta2020Longrange}, by which the position fluctuation is almost completely suppressed.

\begin{figure}[htbp]
\centering
\includegraphics[width=0.9\linewidth]{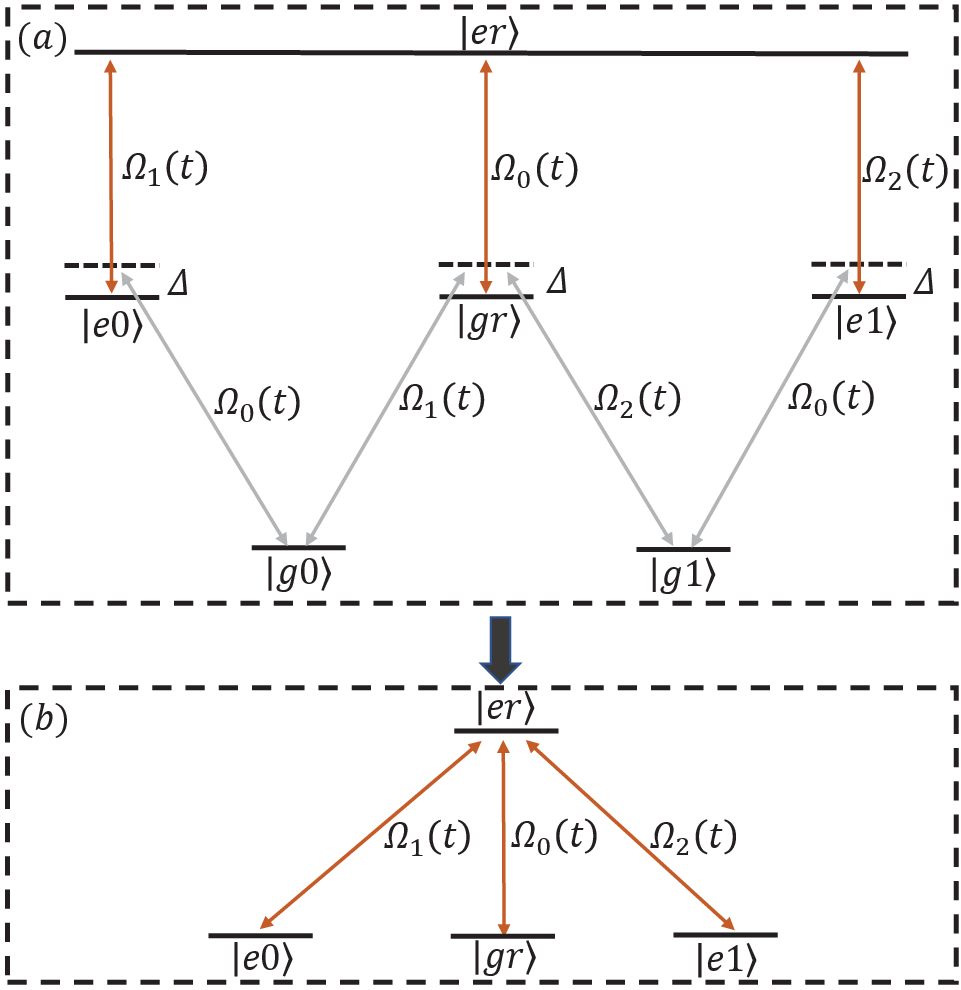}
\caption{(a) Transition diagram for the double-Rydberg-atom system. The transitions plotted with the gray solid lines can be strongly suppressed under the condition of $U_{12}\approx\Delta\gg\Omega_0(t)$, $\Omega_1(t)$, $\Omega_2(t)$. (b) Effective four-level configuration for the double-Rydberg-atom system.}\label{energylevel_twoqubit}
\end{figure}

In the rotating frame with respect to $\mathcal{U}(t)=\exp{(-iU_{12}t|er\rangle\langle er|)}$, the Hamiltonian in Eq.~(\ref{H}) is transformed to
\begin{equation}\label{H_rot}
\begin{aligned}
& H_{\rm rot}(t)\\=& \Omega_0(t)e^{-i\Delta t}
\left(|e0\rangle\langle g0|+|e1\rangle\langle g1|+e^{iU_{12}t}|er\rangle\langle gr|\right)\\
+& \Omega_1(t)e^{-i(\Delta t+\varphi_1)}\left(|gr\rangle\langle g0|+e^{iU_{12}t}|er\rangle\langle e0|\right)\\
+& \Omega_2(t)e^{-i(\Delta t+\varphi_2)}\left(|gr\rangle\langle g1|+e^{iU_{12}t}|er\rangle\langle e1|\right)+{\rm H.c.},
\end{aligned}
\end{equation}
as shown by the transition diagram in Fig.~\ref{energylevel_twoqubit}(a). Under the far-off-detuning condition
$U_{12}\approx\Delta\gg\{\Omega_0(t), \Omega_1(t), \Omega_2(t)\}$, a number of unwanted transitions [see the gray solid lines in Fig.~\ref{energylevel_twoqubit}(a)] are strongly suppressed. As long as $U_{12}$ and $\Delta$ are sufficiently large and close to each other in magnitude, one can always omit the fast-oscillating terms in $H_{\rm rot}(t)$ under the rotating-wave approximation. Then the Hamiltonian in Eq.~(\ref{H_rot}) can be effectively expressed by a four-level configuration
\begin{equation}\label{H_eff}
\begin{aligned}
H_{\rm eff}(t)&=\Omega_0(t)|er\rangle\langle gr|+\Omega_1(t)e^{-i\varphi_1}|er\rangle\langle e0|\\
&+\Omega_2(t)e^{-i\varphi_2}|er\rangle\langle e1|+{\rm H.c.},
\end{aligned}
\end{equation}
as shown in Fig.~\ref{energylevel_twoqubit}(b). When the laser field on the control atom is turned off, i.e., $\Omega_0(t)=0$, Fig.~\ref{energylevel_twoqubit}(b) reduces exactly to the $\Lambda$-level configuration in the standard NHQC scheme~\cite{Sjoqvist2012Nonadiabatic}.

To construct the nonadiabatic holonomic two-qubit controlled gates with dark paths, the driving fields on the target atom can be parametrized with $\Omega_1(t)=\Omega(t)\sin{\theta/2}$ and $\Omega_2(t)=-\Omega(t)\cos{\theta/2}$, where $\Omega(t)$ and $\theta$ are time dependent and time independent, respectively. With the Morris-Shore transformation~\cite{Morris1983Reduction}, the effective Hamiltonian in Eq.~(\ref{H_eff}) can be recast as
\begin{equation}\label{H_effdb}
H_{\rm eff}(t)=\Omega(t)e^{i\varphi_2}|b\rangle\langle er|+\Omega_0(t)|gr\rangle\langle er|+{\rm H.c.}
\end{equation}
in the dark-bright basis with
\begin{equation}\label{dbbasistwo}
\begin{aligned}
|b\rangle&=\sin{\frac{\theta}{2}}e^{i\varphi}|e0\rangle-\cos{\frac{\theta}{2}}|e1\rangle,\\
|D_1\rangle&=\cos{\frac{\theta}{2}}|e0\rangle+\sin{\frac{\theta}{2}}e^{-i\varphi}|e1\rangle,
\end{aligned}
\end{equation}
where $\varphi\equiv\varphi_1-\varphi_2$ and $|D_1\rangle$ is the dark state with zero eigenvalue, which remains time independent during the evolution for invariant $\theta$ and $\varphi$ and forms the first dark path in our scheme.

The second dark path $|D_2\rangle=|D_2(t)\rangle$ is generally time dependent in our scheme, whose ans\"atz could be derived by two constraints: (i) $|D_2\rangle$ is always orthogonal to $|D_1\rangle$, i.e., $\langle D_1|D_2\rangle=0$; and (ii) the expectation value $\langle D_2|H_{\rm eff}(t)|D_2\rangle$ remains vanishing during the gate construction, which accumulates no dynamical phase. Accordingly, $|D_2\rangle$ can be expressed by
\begin{equation}\label{darkpath}
\begin{aligned}
|D_2\rangle&=\cos u(t) \cos v(t) e^{i\varphi_2}|b\rangle-i\sin u(t)|er\rangle \\
&-\cos u(t) \sin v(t)|gr\rangle
\end{aligned}
\end{equation}
with two time-dependent parameters $u(t)$ and $v(t)$. By the Schr\"odinger equation $H_{\rm eff}(t)|D_2(t)\rangle=id|D_2(t)\rangle/dt$, the time-dependent Rabi frequencies $\Omega(t)$ and $\Omega_0(t)$ can be inversely determined by
\begin{equation}\label{Om_Omtildetwo}
\begin{aligned}
\Omega(t)&=\dot{v}(t)\cot{u(t)}\sin{v(t)}+\dot{u}(t)\cos{v(t)},\\
\Omega_0(t)&=\dot{v}(t)\cot{u(t)}\cos{v(t)}-\dot{u}(t)\sin{v(t)}.
\end{aligned}
\end{equation}
Under the cyclic condition for the gate construction lasting a period of $\tau$, i.e., $\{|D_1\rangle, |D_2(\tau)\rangle\}=\{|e0\rangle, |e1\rangle\}$, the boundary conditions of $u(t)$ and $v(t)$ have to be $u(\tau)=u(0)=v(\tau)=v(0)=0$. We follow Refs.~\cite{Ming2022Experimental,Andr2022Dark} and choose
\begin{equation}\label{uv}
u(t)=\frac{\pi}{2}\sin^2\left(\frac{\pi t}{\tau}\right), \quad v(t)=\eta[1-\cos{u(t)}]
\end{equation}
to ensure the cyclic evolution. By Eqs.~(\ref{Om_Omtildetwo}) and (\ref{uv}), the tunable parameter $\eta$ controls both the driving laser applied on the control atom and the transition between $|er\rangle$ and $|gr\rangle$ as shown in Fig.~\ref{energylevel_twoqubit}(b). Specifically, a vanishing $\eta$ indicates $v(t)=0$ in Eq.~(\ref{uv}), and it leads to $\Omega(t)\ne0$ and $\Omega_0(t)=0$ in Eq.~(\ref{Om_Omtildetwo}), which describes the standard NHQC scheme~\cite{Herterich2016Singleloop}. While a nonzero $\eta$ describes our NHGTD scheme.

To implement a universal set of nonadiabatic holonomic two-qubit controlled gates, we adopt the multipulse single-loop method~\cite{Herterich2016Singleloop}, in which the loop for gate construction is divided into two segments with equal intervals. Accordingly, the cyclic evolution is completed by concatenating two unitary operators in the effective subspace:
\begin{equation}\label{unitary_part}
\begin{aligned}
U(\tau/2,0)&=|D_1\rangle\langle D_1|+|D_2(\tau/2)\rangle\langle D_2(0)|\\
&=|D_1\rangle\langle D_1|-i|er\rangle\langle b|,\\
U(\tau,\tau/2)&=|D_1\rangle\langle D_1|+|D_2(\tau)\rangle\langle D_2(\tau/2)|\\
&=|D_1\rangle\langle D_1|+ie^{i\gamma}|b\rangle\langle er|,
\end{aligned}
\end{equation}
where the phase shift $\gamma$ appears to be $\varphi_2$ in the second path segment. Consequently, the holonomic matrix for the whole geometric evolution is given by
\begin{equation}\label{Utau}
U(\tau,0)=|D_1\rangle\langle D_1|+e^{i\gamma}|b\rangle\langle b|.
\end{equation}
In the gate space spanned by $\{|e0\rangle, |e1\rangle\}$, it can be rewritten as
\begin{equation}\label{unitary}
U(\theta, \varphi, \gamma)=U(\tau,0)=|e\rangle\langle e|\otimes e^{i(\gamma/2)}e^{-i(\gamma/2)\vec{n}\cdot\vec{\sigma}},
\end{equation}
where $\vec{n}\equiv(\sin{\theta}\cos{\varphi},\sin{\theta}\sin{\varphi},\cos{\theta})$, $\vec{\sigma}$ is the Pauli matrix, and $\exp{(i\gamma/2)}$ is a global phase factor. The unitary matrix $U(\theta,\varphi,\gamma)$ in Eq.~(\ref{unitary}) allows the target atom to rotate about the desired axis $\vec{n}$ by an arbitrary angle $\gamma$, provided the control atom is prepared at the Rydberg state $|e\rangle$. The parameters $\theta$, $\varphi$, and $\gamma$ can be controlled by the Rabi frequencies and phases of the driving lasers. Therefore, $U(\theta,\varphi,\gamma)$ constructs an arbitrary nonadiabatic holonomic two-qubit controlled gate.

\subsection{Gate performance}

The performance of any quantum gates should be practically measured by the gate fidelity subject to external noises and systematic errors. The former could be described by the master equation~\cite{Carmichael1999statistical},
\begin{equation}\label{master_2qubit}
\begin{aligned}
\frac{\partial \rho}{\partial t}&=-i[H(t),\rho]+\frac{\kappa}{2}\mathcal{L}(\sigma^{-}_{c})+\frac{\kappa_z}{2}\mathcal{L}(\sigma^{z}_{c})\\
&+\frac{\kappa_0}{2}\mathcal{L}(|0\rangle_t\langle r|)+\frac{\kappa_1}{2}\mathcal{L}(|1\rangle_t\langle r|)+\frac{\kappa_z}{2}\mathcal{L}(\sigma^z_{t}).
\end{aligned}
\end{equation}
Here $\rho$ is the density matrix for the coupled Rydberg atoms, $H(t)$ is their full Hamiltonian in Eq.~(\ref{H}), and $\mathcal{L}(o)$ is the Lindblad superoperator defined as $\mathcal{L}(o)\equiv2o\rho o^\dagger-o^\dagger o\rho-\rho o^\dagger o$~\cite{Scully1997quantum}, where $o=\sigma^-_{c}, \sigma^z_{c}, |0\rangle_t\langle r|, |1\rangle_t\langle r|, \sigma^z_{t}$. The superscripts and subscripts of these operators indicate the type of quantum channels and the atom under decoherence, respectively. For example, $\sigma^-_{c}\equiv|g\rangle_{c}\langle e|$ represents the dissipation channel of the control atom with a decay rate $\kappa$, and $|0\rangle_t\langle r|$ and $|1\rangle_t\langle r|$ represent the spontaneous emission of the target atom from the Rydberg state $|r\rangle$ to the ground states $|0\rangle$ and $|1\rangle$ with decay rates $\kappa_0$ and $\kappa_1$, respectively. For simplicity, we assume $\kappa_0=\kappa_1=\kappa/2$. The dephasing processes of the control atom and the target atom are described by $\sigma^z_{c}\equiv|g\rangle_{c}\langle g|-|e\rangle_{c}\langle e|$ and $\sigma^z_{t}\equiv|0\rangle_t\langle 0|+|1\rangle_t\langle 1|-|r\rangle_t\langle r|$, respectively, with the same rate $\kappa_z$. The decoherence rate $\kappa$ of Rydberg states of the alkali-metal atoms~\cite{Beterov2009Quasiclassical,Isenhower2010Demonstration,Adams2020Rydberg} with a low angular momentum relates to the quantum numbers $n$ and $l$~\cite{Adams2020Rydberg}, which are typically in the range of $0.75-2.48$ kHz~\cite{Beterov2009Quasiclassical,Isenhower2010Demonstration,Sun2021Onestep,Sun2022Onestep} around room temperature; $\kappa_z$ is close to $\kappa$ in magnitude~\cite{Sun2021Onestep}. In our simulation, $\kappa$ is set in the order of kilohertz~\cite{Sun2021Onestep,Sun2022Onestep,Sun2022Onestep} and the dephasing channel indicated by $\kappa_z=2\pi\times1$ kHz is always on.

The systematic errors in Rabi frequencies (driving strengths) could be grouped into global and local types, representing the deviation for the full system and for the target atom, respectively. Practically, the Hamiltonian $H(t)$ in Eq.~(\ref{master_2qubit}) becomes
\begin{equation}\label{H_error}
H(t)=(1+\epsilon)[H_c(t)+(1+\alpha)H_t(t)]+U_{12}|er\rangle\langle er|,
\end{equation}
where $\epsilon$ and $\alpha$ are the dimensionless coefficients for the global and local Rabi-frequency errors, respectively.

To evaluate the gate performance of our NHGTD scheme, we adopt the average fidelity function defined in Refs.~\cite{Monz2009Realization,Liang2023Stateindependent},
\begin{equation}\label{fidelity}
F=\frac{1}{N}\sum_{n=1}^{N}\langle\psi_n(0)|U^\dagger\rho(\tau)U|\psi_n(0)\rangle.
\end{equation}
Here $\rho(\tau)$ is obtained by Eqs.~(\ref{master_2qubit}) and (\ref{H_error}) with the $n$th benchmark state $|\psi_n(0)\rangle=|\psi_{c}(0)\rangle\otimes|\psi_t(0)\rangle$ and the fixed operation time $\tau$ for quantum gate. In addition, $U=U(\theta, \varphi, \gamma)$ is the ideal two-qubit controlled gate (holonomic transformation) given in Eq.~(\ref{unitary}). In numerical simulation, the initial states of the control atom $|\psi_{c}(0)\rangle$ and the target atom $|\psi_t(0)\rangle$ are sampled from the state sets $\{|g\rangle$, $|e\rangle$, $(|g\rangle+|e\rangle)/\sqrt{2}$, $(|g\rangle-|e\rangle)/\sqrt{2}$, $(|g\rangle+i|e\rangle)/\sqrt{2}$, $(|g\rangle-i|e\rangle)/\sqrt{2}\}$ and $\{|0\rangle$, $|1\rangle$, $(|0\rangle+|1\rangle)/\sqrt{2}$, $(|0\rangle-|1\rangle)/\sqrt{2}$, $(|0\rangle+i|1\rangle)/\sqrt{2}$, $(|0\rangle-i|1\rangle)/\sqrt{2}\}$, respectively. Thus $N=6^2=36$ states are used in testing the two-qubit controlled gates. The choice of benchmark states does not have a significant influence on the gate performance. The type of controlled gate and the parameters $\theta$, $\varphi$ and $\gamma$ are determined by the amplitudes and phases of the driving fields. We focus here on the CNOT gate $U(\pi/2,0,\pi)$ and the CZ gate $U(0,0,\pi)$.

\begin{figure}[htbp]
\centering
\includegraphics[width=0.9\linewidth]{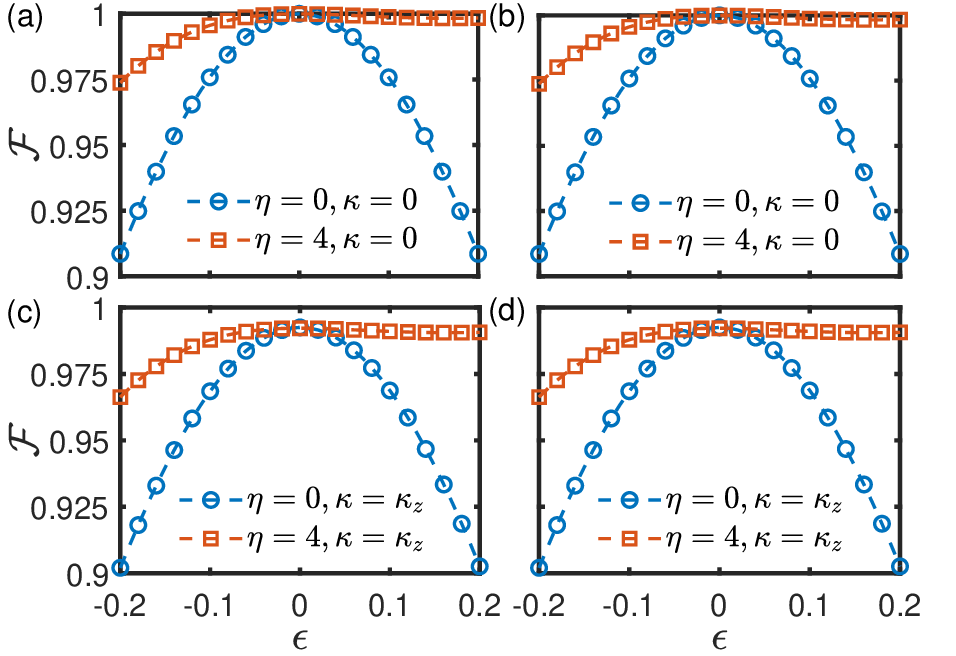}
\caption{Comparison of the average fidelity of our NHGTD (with $\eta=4$) and the existing NHQC (with $\eta=0$) schemes under the global error $\epsilon$ in Rabi frequency of driving fields for (a) and (c) CNOT gates and (b) and (d) CZ gates in the (a) and (b) absence and (c) and (d) presence of dissipation. The detuning and the Rydberg-like interaction strength are $\Delta=U_{12}=5\times10^{5}\kappa_z$ and in (c) and (d) the decay rate is set as $\kappa=\kappa_z$, where the dephasing rate is $\kappa_z=2\pi\times1$ kHz. The Rabi frequencies defined in Eq.~(\ref{Om_Omtildetwo}) are not greater than $3$ MHz. The local error in Rabi frequency is set as $\alpha=0$. }\label{errortwo}
\end{figure}

In Fig.~\ref{errortwo} we compare our NHGTD scheme and the standard NHQC scheme in the gate fidelities against the global error in Rabi frequency. It demonstrates that in the presence or in the absence of the dissipation channels of the Rydberg atoms, the driving field on the control atom ($\eta=4$) enhances significantly the gate robustness against the global Rabi-frequency error of the NHQC scheme ($\eta=0$) for both CNOT and CZ gates in the whole range of $\epsilon\in[-0.2, 0.2]$. More interesting is that the fidelity of our dark-path scheme is almost insensitive to a positive $\epsilon$, which makes an asymmetric performance of our gates around $\epsilon=0$. It is reasonable because a positive deviation means an enhanced driving laser that will reduce the running time of the holonomic gates.

For the CNOT gate with no dissipation in Fig.~\ref{errortwo}(a), its fidelity can be maintained over $F=0.97$ with $\eta=4$, higher than $F=0.91$ with $\eta=0$, even when the relative global error is as large as $\epsilon=-0.2$. For both CNOT and CZ gates [see Figs.~\ref{errortwo}(a) and \ref{errortwo}(b)] with $\epsilon\geq0$, our fidelity remains almost unit. Turning on the dissipation channel does not violate the advantage of our NHGTD scheme over the NHQC scheme, although it slightly reduces the fidelities for both CNOT and CZ gates. Note that, around the zero global error about the Rabi frequency, the previous dark-path scheme in the trapped-ion system~\cite{Ming2022Experimental,Andr2022Dark} has a worse performance than the standard NHQC scheme. In contrast, we can find that in both Figs.~\ref{errortwo}(c) and \ref{errortwo}(d) the fidelity $F=0.99$ with driving on the control atom $\eta=4$ prevails over $F=0.98$ with no driving $\eta=0$. This is due to the fact that our dark-path scheme in Rydberg atoms is conveniently controlled by an off-resonant laser field on the control atom without coupling the upper level of the target atom to an ancillary level accompanied by unwanted leakage.

\begin{figure}[htbp]
\centering
\includegraphics[width=0.9\linewidth]{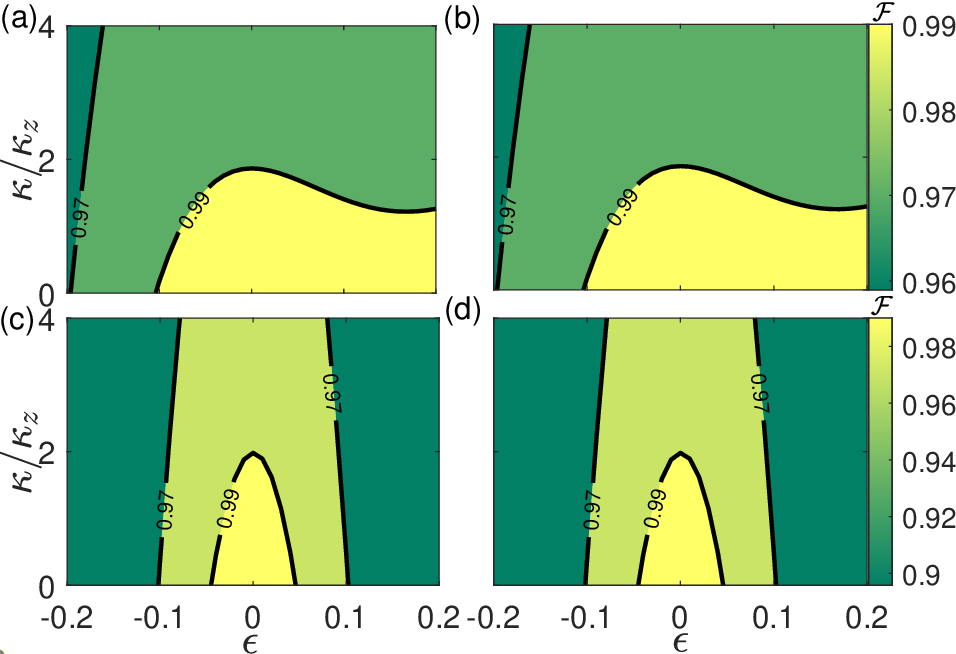}
\caption{Fidelity landscapes of NHGTD [with $\eta=4$ in (a) and (b)] and NHQC [with $\eta=0$ in (c) and (d)] schemes in the space of global error $\epsilon$ in Rabi frequency of driving fields and decay rate $\kappa$ for (a) and (c) CNOT gates and (b) and (d) CZ gates. The local Rabi frequency error is set as $\alpha=0$. The other parameters are the same as those in Fig.~\ref{errortwo}.}\label{two_epsilon}
\end{figure}

The advantage of our NHGTD scheme over NHQC schemes can be extended to the whole space of the global error coefficient $\epsilon$ and the decay rate $\kappa$ as shown in Fig.~\ref{two_epsilon}. The fidelity landscape is divided into the regimes of $F\le0.97$ (dark green area), $0.97\le F\le0.99$ (light green area), and $F\ge0.99$ (yellow area). For either the CNOT or CZ gate, our scheme tolerates a much stronger decoherence in the presence of a positive global error. In particular, for $\epsilon=\pm0.04$, a CNOT gate attains $F=0.99$ even if $\kappa=1.7\kappa_z$ within our scheme. In contrast, the same high-fidelity gate survives when $\kappa\leq0.2\kappa_z$ within the NHQC scheme.

\begin{figure}[htbp]
\centering
\includegraphics[width=0.9\linewidth]{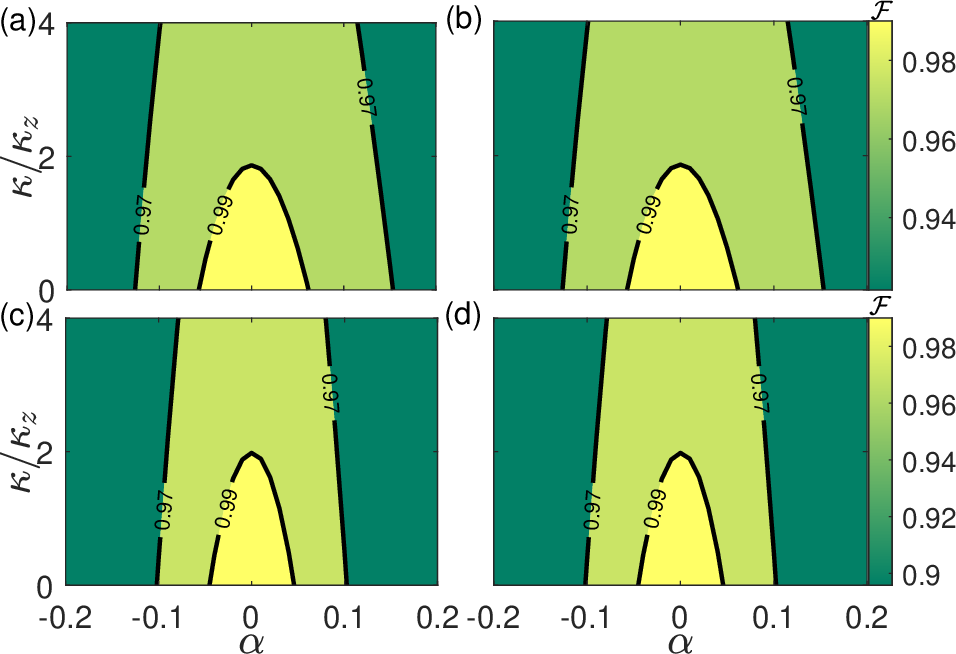}
\caption{Fidelity landscapes of our NHGTD [with $\eta=4$ in (a) and (b)] and the NHQC [with $\eta=0$ in (c) and (d)] schemes in the space of local error $\alpha$ in Rabi frequency of driving fields and decay rate $\kappa$ for (a) and (c) CNOT gates and (b) and (d) CZ gates. The global Rabi frequency error is set as $\epsilon=0$. The other parameters are the same as those in Fig.~\ref{errortwo}.}\label{decaytwo}
\end{figure}

Similarly, Fig.~\ref{decaytwo} demonstrates the fidelity landscape for both CNOT and CZ gates in the parameter space of local error coefficient $\alpha$ and decay rate $\kappa$. Note that the high-fidelity regimes of our scheme ($\eta=4$) are much broader than those of the NHQC scheme ($\eta=0$) as well. In particular, when $\kappa=\kappa_z$, the CNOT-gate fidelity of our scheme can be maintained as $F=0.99$ under the local Rabi-frequency error $\alpha=\pm0.05$. In contrast, the same performance can survive up to $\alpha=\pm0.03$ in the NHQC scheme. For $\alpha=\pm0.2$, the CNOT-gate fidelity $F=0.99$ is preserved when $\kappa=1.1\kappa_z$ within our scheme, while $\kappa=0.4\kappa_z$ within the NHQC scheme. The off-resonant laser field on the control atom can thus improve the gate robustness in resisting both global and local errors in Rabi frequency.

\section{Nonadiabatic holonomic three-qubit controlled gates}\label{modelandHam_three}

\subsection{Gate construction with dark paths}

\begin{figure}[htbp]
\centering
\includegraphics[width=0.85\linewidth]{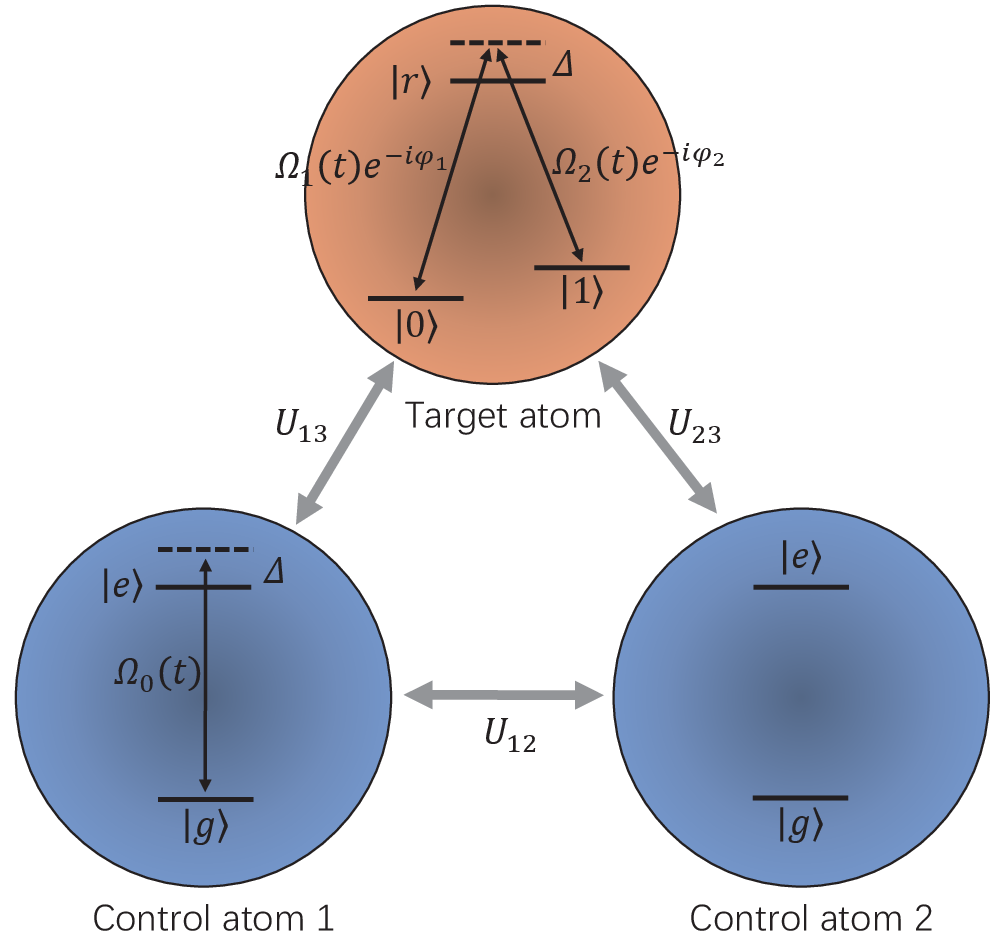}
\caption{Sketch of three coupled Rydberg atoms, where the first control atom and the target atom are under off-resonant driving. Here $U_{ij}$ ($i<j$) is the Rydberg-mediated interaction between the Rydberg states of atoms.}\label{model1}
\end{figure}

The three-qubit controlled gates can be constructed with a system of three coupled Rydberg atoms as shown in Fig.~\ref{model1}. Again the control atoms and the target atom are considered as two-level and three-level systems, respectively. Note that the second control atom is not under driving, which is merely coupled to the target atom and the first control atom via the Rydberg-mediated interactions $U_{23}$ and $U_{12}$, respectively.

In the interaction picture with respect to the free Hamiltonian of the atoms, the full Hamiltonian reads
\begin{equation}\label{H1}
H(t)=H_{c_1}(t)+H_t(t)+H_I,
\end{equation}
where
\begin{equation}\label{H1single}
\begin{aligned}
H_{c_1}(t)&=\Omega_0(t)e^{-i\Delta t}|e\rangle_{c_1}\langle g|+{\rm H.c.},\\
H_t(t)&=\Omega_1(t)e^{-i(\Delta t+\varphi_1)}|r\rangle_t\langle 0|\\ &+\Omega_2(t)e^{-i(\Delta t+\varphi_2)}|r\rangle_t\langle 1|+{\rm H.c.}
\end{aligned}
\end{equation}
are the driving Hamiltonians for the first control atom and the target atom, respectively, and
\begin{equation}\label{H1I}
\begin{aligned}
H_I&=U_{13}|e\rangle_{c_1}\langle e|\otimes\mathcal{I}_{c_2}\otimes|r\rangle_t\langle r|+U_{23}\mathcal{I}_{c_1}\otimes|er\rangle_{c_2t}\langle er|\\ &+U_{12}|ee\rangle_{c_1c_2}\langle ee|\otimes\mathcal{I}_t
\end{aligned}
\end{equation}
describes the interactions among the three atoms. In the optical lattice~\cite{Zhang2017coherent,Graham2019Rydberg,Gambetta2020Engineering,Gambetta2020Longrange,Yin2020Onestep,Sun2021Onestep} where the Rydberg atoms are strongly confined ($\delta d_{ij}\approx0$) within a harmonic potential via the phonon mode, the atomic interaction consists of the bare Rydberg-like interaction and the effective Rydberg-like interaction induced by the electron-phonon coupling, both of which depend on the atomic distance. Our scheme works with an interaction strength similar to that in the optical lattice~\cite{Gambetta2020Engineering}, e.g., $U_{ij}\sim500$ MHz when $d_{ij}\sim5.3 \mu$m. The three driving intensities $\Omega_j$, $j=0,1,2$, are much smaller than $U_{ij}$. In contrast, the conventional NHQC scheme~\cite{Sun2021Onestep} for the three-qubit gates demands that $U_{12}$ is in the strong-coupling regime (scaling with $500$ MHz) and $U_{13}$ and $U_{23}$ are even in the ultrastrong-coupling regime (scaling with $3600$ MHz). In addition, the intensities of the driving fields on the target atom, $\Omega_1$ and $\Omega_2$, have to be much smaller than that on the control atom, $\Omega_0$, limiting the gate speed.

\begin{figure}[htbp]
\centering
\includegraphics[width=0.9\linewidth]{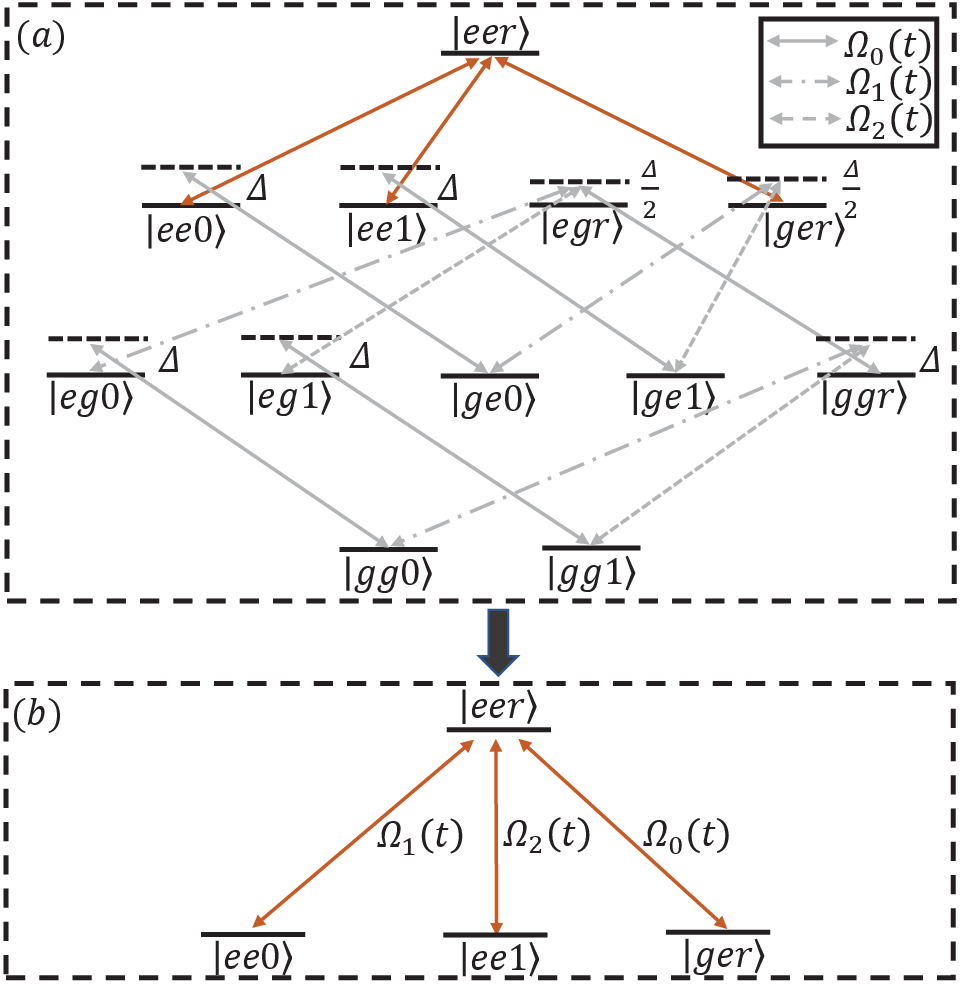}
\caption{(a) Transition diagram for the three-Rydberg-atom system. The transitions plotted with the gray lines can be strongly suppressed when $\Delta\approx U_{12}+U_{13}\approx U_{13}+U_{23}$ and $ U_{ij}\gg\{\Omega_0(t), \Omega_1(t), \Omega_2(t)\}$ ($i<j$). (b) Effective four-level configuration for the three-Rydberg-atom system.}\label{energylevel_threequbit}
\end{figure}

In the rotating frame with respect to $\mathcal{U}=\exp(-iH_It)$, the Hamiltonian in Eq.~(\ref{H1}) can be rewritten as
\begin{equation}\label{H1_rot}
H_{\rm rot}(t)=H'_{c_1}(t)+H'_t(t),
\end{equation}
where
\begin{equation}
\begin{aligned}
H'_{c_1}(t)&=\Omega_0(t)e^{-i\Delta t}(|eg0\rangle\langle gg0|+|eg1\rangle\langle gg1|\\
&+e^{iU_{13}t}|egr\rangle\langle ggr|+e^{iU_{12}t}|ee0\rangle\langle ge0|\\
&+e^{iU_{12}t}|ee1\rangle\langle ge1|+e^{i(U_{12}+U_{13})t}|eer\rangle\langle ger|)\\
&+{\rm H.c.},
\end{aligned}
\end{equation}
and
\begin{equation}
\begin{aligned}
H'_t(t)&=\Omega_1(t)e^{-i(\Delta t+\varphi_1)}(|ggr\rangle\langle gg0|+e^{iU_{23}t}|ger\rangle\langle ge0|\\
&+e^{iU_{13}t}|egr\rangle\langle eg0|+e^{i(U_{13}+U_{23})t}|eer\rangle\langle ee0|)\\
&+\Omega_2(t)e^{-i(\Delta t+\varphi_2)}(|ggr\rangle\langle gg1|+e^{iU_{23}t}|ger\rangle\langle ge1|)\\
&+e^{iU_{13}t}|egr\rangle\langle eg1|+e^{i(U_{13}+U_{23})t}|eer\rangle\langle ee1|\\
&+{\rm H.c.},
\end{aligned}
\end{equation}
as shown by the transition diagram in Fig.~\ref{energylevel_threequbit}(a). Under the far-off-detuning condition $\Delta\approx U_{12}+U_{13}\approx U_{13}+U_{23}$ and $U_{ij}\gg\Omega_0(t), \Omega_1(t), \Omega_2(t)$ ($i<j$), unwanted transitions [see the gray lines in Fig.~\ref{energylevel_threequbit}(a)] are strongly suppressed. Note that $U_{13}$ is not necessarily the same as $U_{12}$ and $U_{23}$. In Fig.~\ref{energylevel_threequbit}(a), the gray solid lines represent the off-resonant terms with Rabi frequency $\Omega_0(t)$ in $H'_{c_1}(t)$, and the gray dashed lines and gray dot-dashed lines represent the off-resonant terms with Rabi frequencies $\Omega_1(t)$ and $\Omega_2(t)$ in $H'_t(t)$, respectively. The survival transitions constitute an effective Hamiltonian again with a four-level configuration
\begin{equation}\label{H1_eff}
\begin{aligned}
H_{\rm eff}(t)&=\Omega_0(t)|eer\rangle\langle ger|+\Omega_1(t)e^{-i\varphi_1}|eer\rangle\langle ee0|\\
&+\Omega_2(t)e^{-i\varphi_2}|eer\rangle\langle ee1|+{\rm H.c.},
\end{aligned}
\end{equation}
as shown in Fig.~\ref{energylevel_threequbit}(b). Similar to the double-atom system, Fig.~\ref{energylevel_threequbit}(b) reduces exactly to the standard NHQC scheme when the driving field on the first control atom is turned off.

Using the same parametric setting as in Eq.~(\ref{H_effdb}) and the Morris-Shore transformation, the effective Hamiltonian in Eq.~(\ref{H1_eff}) can be written as
\begin{equation}\label{H1_effdb}
H_{\rm eff}(t)=\Omega(t)e^{i\varphi_2}|b\rangle\langle eer|+\Omega_0(t)|ger\rangle\langle eer|+{\rm H.c.},
\end{equation}
where the bright state and the first dark path are
\begin{equation}\label{dbbasisthree}
\begin{aligned}
|b\rangle&=\sin{\frac{\theta}{2}}e^{i\varphi}|ee0\rangle-\cos{\frac{\theta}{2}}|ee1\rangle, \\
|D_1\rangle&=\cos{\frac{\theta}{2}}|ee0\rangle+\sin{\frac{\theta}{2}}e^{-i\varphi}|ee1\rangle,
\end{aligned}
\end{equation}
respectively. Under the similar constraints as those in the double-qubit-gate case, the second dark path can be given by
\begin{equation}\label{darkpath_3qubit}
\begin{aligned}
|D_2(t)\rangle&=\cos{u(t)}\cos{v(t)}e^{i\varphi_2}|b\rangle-i\sin{u(t)}|eer\rangle,\\
&-\cos{u(t)}\sin{v(t)}|ger\rangle,
\end{aligned}
\end{equation}
where the two time-dependent parameters $u(t)$ and $v(t)$ could also be chosen as in Eq.~(\ref{uv}) due to the boundary condition. The time-dependent parameters $\Omega(t)$ and $\Omega_0(t)$ still follow the same relation given by Eq.~(\ref{Om_Omtildetwo}) with $u(t)$ and $v(t)$.

The multipulse single-loop method~\cite{Herterich2016Singleloop} adapts to construct a universal set of nonadiabatic holonomic three-qubit controlled gates, by which the cyclic evolution loop $U(\tau, 0)$ is a product of
\begin{equation}\label{U3qubit}
\begin{aligned}
U\left(\frac{\tau}{2},0\right)&=|D_1\rangle\langle D_1|-i|eer\rangle\langle b|,\\
U\left(\tau,\frac{\tau}{2}\right)&=|D_1\rangle\langle D_1|+ie^{i\gamma}|b\rangle\langle eer|,
\end{aligned}
\end{equation}
resulting in the same form of Eq.~(\ref{Utau}). In the gate space spanned by $\{|ee0\rangle, |ee1\rangle\}$, the unitary matrix for an arbitrary nonadiabatic holonomic three-qubit controlled gate can be written as
\begin{equation}\label{U3qubitCombine}
U(\theta,\varphi,\gamma)=U(\tau,0)=|ee\rangle\langle ee|\otimes e^{i(\gamma/2)}e^{-i(\gamma/2)\vec{n}\cdot\vec{\sigma}}.
\end{equation}
The target atom can therefore rotate around the desired axis $\vec{n}$ by a desired angle $\gamma$ as long as both control atoms are simultaneously at the high-lying Rydberg state. We focus here on the controlled-controlled-NOT gate by choosing $\theta=\pi/2$, $\varphi=0$, and $\gamma=\pi$.

\subsection{Gate performance}

\begin{figure}[htbp]
\centering
\includegraphics[width=0.9\linewidth]{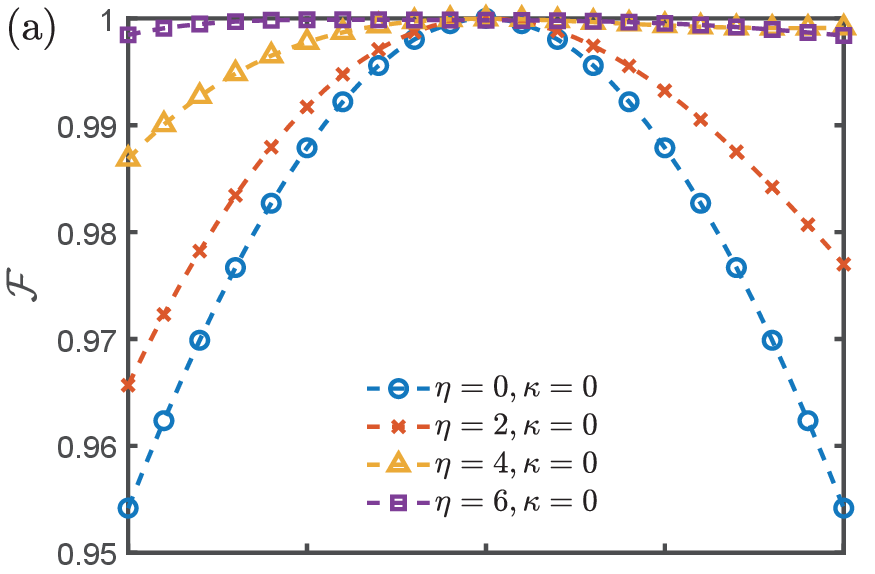}
\includegraphics[width=0.9\linewidth]{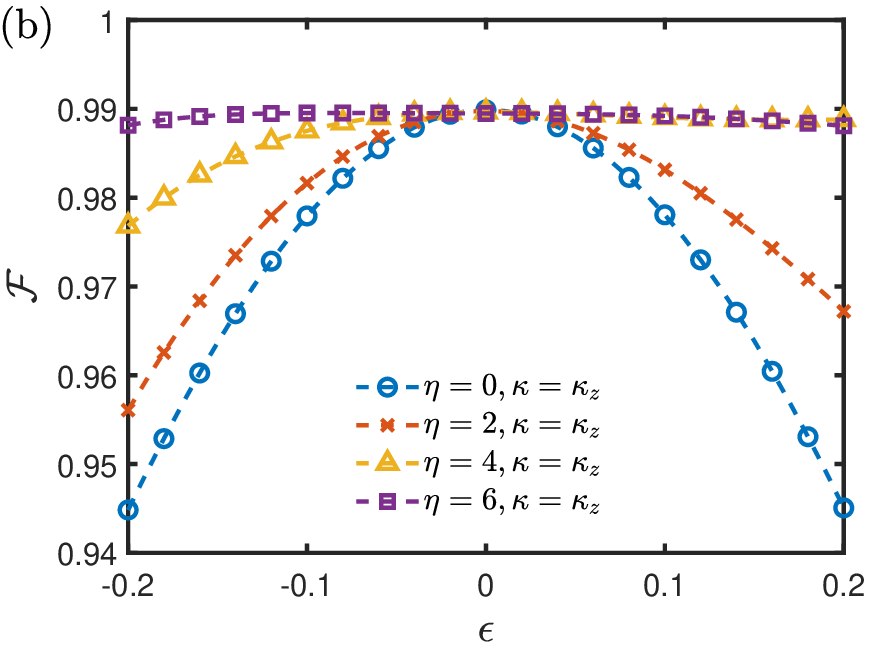}
\caption{Comparisons of the CCNOT-gate fidelity for NHGTD (with nonvanishing $\eta$) and NHQC (with $\eta=0$) schemes under the global error $\epsilon$ in Rabi frequency of the driving fields for (a) $\kappa=0$ (dissipation-free) and (b) $\kappa=\kappa_z$, with $\Delta/2=U_{13}=U_{23}=U_{12}=5\times10^5\kappa_z$ and $\kappa_z=2\pi\times1$ kHz. The other parameters are the same as in Fig.~\ref{errortwo}.}\label{errorthree}
\end{figure}

Taking the second control atom into consideration, the master equation (\ref{master_2qubit}) becomes
\begin{equation}\label{master_3qubit}
\begin{aligned}
\frac{\partial \rho}{\partial t}&=-i[H(t),\rho]+\sum_{i=1}^{2}\left[\frac{\kappa}{2}\mathcal{L}(\sigma_{c_i}^{-})
+\frac{\kappa_z}{2}\mathcal{L}(\sigma_{c_i}^{z})\right]\\
&+\frac{\kappa}{4}\mathcal{L}(|0\rangle_t\langle r|)+\frac{\kappa}{4}\mathcal{L}(|1\rangle_t\langle r|)+\frac{\kappa_z}{2}\mathcal{L}(\sigma_t^z),
\end{aligned}
\end{equation}
where $H(t)$ is the full Hamiltonian of three coupled Rydberg atoms in Eq.~(\ref{H1}). With the systematic errors in the driving strengths (Rabi frequencies), the Hamiltonian in Eq.~(\ref{H1}) becomes
\begin{equation}\label{H1_error}
H(t)=(1+\epsilon)[H_{c_1}(t)+(1+\alpha)H_t(t)]+H_I,
\end{equation}
where $\epsilon$ and $\alpha$ represent again the dimensionless coefficients for the global and local errors, respectively. We use the average fidelity defined in Eq.~(\ref{fidelity}) to evaluate the performance of our NHGTD scheme in constructing the three-qubit gates, where the benchmark states for the control atoms are chosen from the set of $\{|g\rangle, |e\rangle, (|g\rangle+|e\rangle)/\sqrt{2}, (|g\rangle-|e\rangle)/\sqrt{2}\}$ and those for the target atom from $\{|0\rangle, |1\rangle, (|0\rangle+|1\rangle)/\sqrt{2}, (|0\rangle-|1\rangle)/\sqrt{2}\}$. Therefore, $N=4^3=64$ states therefore used in the simulation.

\begin{figure}[htbp]
\centering
\includegraphics[width=0.9\linewidth]{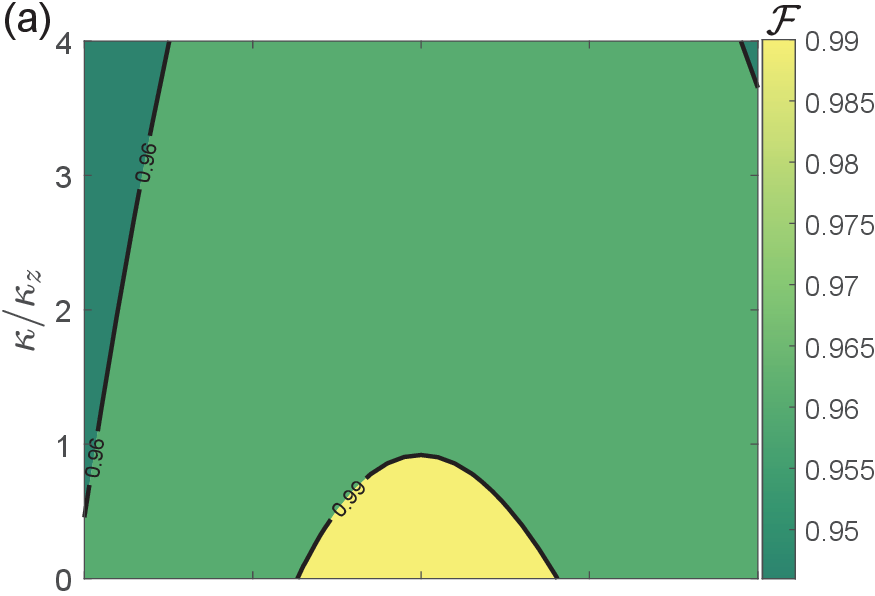}
\includegraphics[width=0.9\linewidth]{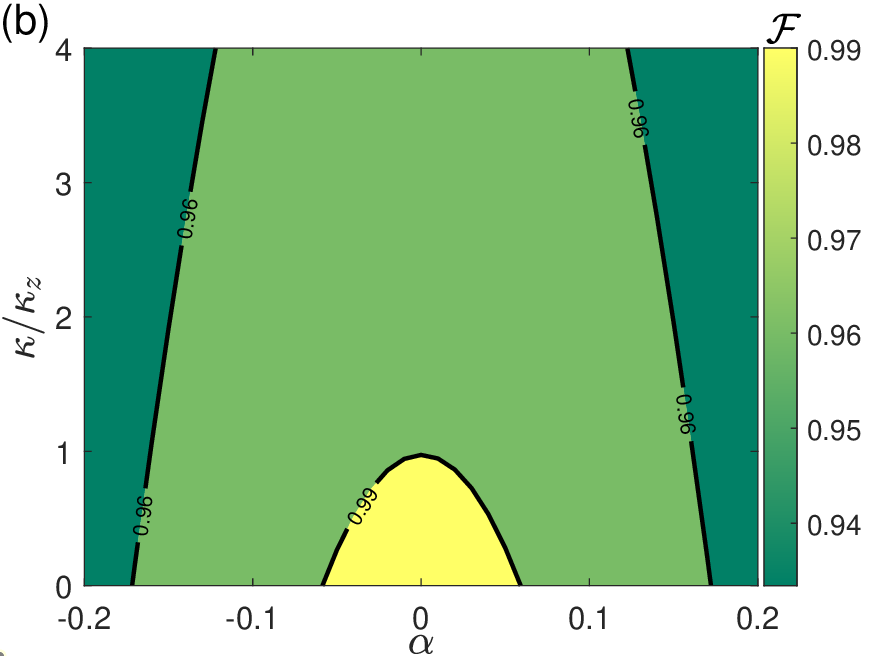}
\caption{Comparisons of the CCNOT-gate fidelity landscape for NHGTD [with $\eta=4$ in (a)] and NHQC [with $\eta=0$ in (b)] schemes in the space of local error $\alpha$ in Rabi frequency of driving fields and decay rate $\kappa$. The global error is set as $\epsilon=0$. The other parameters are the same as in Fig.~\ref{errorthree}.}\label{decaythree}
\end{figure}

In Fig.~\ref{errorthree} we compare our scheme and the standard NHQC scheme in the CCNOT gate fidelities under the global error in Rabi frequency. In the absence [see Fig.~\ref{errorthree}(a)] and in the presence [see Fig.~\ref{errorthree}(b)] of quantum dissipation, it is demonstrated that our NHGTD scheme prevails over the NHQC scheme. It is deterministic that enhancing the driving field on the first control atom can improve significantly the capacity of our CCNOT gate in resisting the global Rabi-frequency error in the whole range of $\epsilon\in[-0.2, 0.2]$. As for the dissipation-free CCNOT gate in Fig.~\ref{errorthree}(a), the fidelities in our scheme can be maintained as high as $F\approx0.97$ and $F\approx0.99$, and almost unit with $\eta=2, 4$ and $6$, respectively, even when $\epsilon=-0.2$. In contrast, the fidelity with $\eta=0$ (NHQC) is less than $0.96$. Note that a high-fidelity gate still favors a positive $\epsilon$ instead of a negative $\epsilon$. In the presence of dissipation in Fig.~\ref{errorthree}(b), the fidelity of the CCNOT gate will drop about $1\%$.

One can find that the CCNOT gate has a better performance than the CNOT gate by comparing Fig.~\ref{errorthree}(a) to Fig.~\ref{errortwo}(a) or by comparing Fig.~\ref{errorthree}(b) to Fig.~\ref{errortwo}(c). Under the same driving of $\eta=4$ and when $\epsilon=-0.2$, the fidelity of the CCNOT gate is higher than that of the CNOT gate by almost $1\%$. This indicates that the holonomic three-qubit gate of our scheme is less susceptible to the global error in Rabi frequency than the double-qubit gate.

The response of our CCNOT gate to the local error in the Rabi frequency can be reflected in the comparison of the landscape in Fig.~\ref{decaythree}(a) for our NHGTD scheme and that in Fig.~\ref{decaythree}(b) for the standard NHQC scheme. It is found that the high-fidelity regimes in our NHGTD scheme are much wider than those in the NHQC scheme. When $\kappa=0.5\kappa_z$, we have $F\geq0.99$ in the range of $\alpha=[-0.05, 0.05]$ in Fig.~\ref{decaythree}(a). In contrast, when $\kappa=0.5\kappa_z$, the fidelity could be maintained about $F\geq0.99$ in the range of $\alpha=[-0.04, 0.04]$ in Fig.~\ref{decaythree}(b).

Additionally, by comparing Fig.~\ref{decaythree}(a) to Fig.~\ref{decaytwo}(a), one can find that the CCNOT gate is more resistant to the local Rabi frequency error $\alpha$ than the CNOT gate, yet it is more susceptible to the environmental dissipation $\kappa\neq0$ in the presence of a small $\alpha$. The bright yellow regime for $F\geq0.99$ in Fig.~\ref{decaythree}(a) is much wider than that in Fig.~\ref{decaytwo}(a). In addition, the upper bound of $\kappa/\kappa_z$ around $\alpha\approx0$ for $F\geq0.99$ is nearly $2.0$ in Fig.~\ref{decaytwo}(a), which is about two times that in Fig.~\ref{decaythree}(a).

\section{Nonadiabatic holonomic $N$-qubit controlled gates}\label{Nqubit}

\begin{figure}[htbp]
\centering
\includegraphics[width=0.85\linewidth]{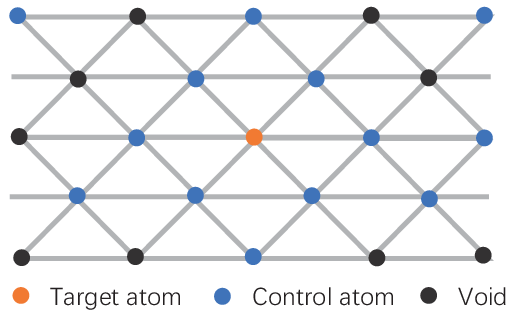}
\caption{Sketch of $N$ Rydberg atoms trapped in a two-dimensional lattice. The orange, blue, and black circles represent the target atom, control atom, and position with no trapped atoms, respectively. }\label{modelN}
\end{figure}

To generalize our NHGTD scheme to the $N$-qubit case, one can consider $N$ Rydberg atoms confined in a two-dimensional optical lattice~\cite{Graham2019Rydberg,Zhang2017coherent,Sun2021Onestep}, as shown in Fig.~\ref{modelN}. Still, only one of the control atoms (the first one) and the target atom (the $N$th one) are under driving, i.e., the number of driving fields does not scale with $N$ and remains as $3$. The control atoms (in a two-level configuration) are coupled to the target atom (in a three-level configuration) via the Rydberg-mediated interaction $U_{iN}$ ($i<N$), and the interactions among the control atoms are $U_{ij}$ ($i<j<N$).

In the interaction picture with respect to the free Hamiltonian of all these Rydberg atoms, the full Hamiltonian of the system has the same form as Eq.~(\ref{H1}), with the driving Hamiltonians for the first control atom and the target atom defined in (\ref{H1single}) and
\begin{equation}\label{HsingleN}
\begin{aligned}
H_I&=\sum_{i=1}^{N-1}U_{iN}|er\rangle_{c_it}\langle er|+\sum_{i<j<N}U_{ij}|er\rangle_{c_ic_j}\langle er|.
\end{aligned}
\end{equation}
In particular, the term $U_{iN}|er\rangle_{c_it}\langle er|$ represents the Rydberg-like interaction between the $i$th control atom and the target atom, and the term $U_{ij}|er\rangle_{c_ic_j}\langle er|$ represents the interaction between the $i$th control atom and the $j$th control atom. In the rotating frame with respect to $\mathcal{U}=\exp(-iH_It)$, the Hamiltonian in Eq.~(\ref{H1}) can be written as
\begin{equation}\label{HrotN}
\begin{aligned}
H_{\rm rot}(t)&=H'_{c_1}(t)+H'_t(t),\\
\end{aligned}
\end{equation}
where
\begin{equation}
\begin{aligned}
H'_{c_1}(t)&=\Omega_0(t)e^{-i\Delta t}(|eg\cdots g0\rangle\langle gg\cdots g0|\\
&+|eg\cdots g1\rangle\langle gg\cdots g1|+e^{iU_{1N}t}|eg\cdots gr\rangle\langle gg\cdots gr|\\
&+\cdots+\left[\exp{i\sum_{j=2}^{N-1}U_{1j}t}\right](|ee\cdots e0\rangle\langle ge\cdots e0|\\
&+|ee\cdots e1\rangle\langle ge\cdots e1|\\
&+e^{iU_{1N}t}|ee\cdots er\rangle\langle ge\cdots er|))+{\rm H.c.}\\
\end{aligned}
\end{equation}
and
\begin{equation}
\begin{aligned}
H'_t(t)&=\Omega_1(t)e^{-i(\Delta t+\varphi_1)}(|gg\cdots gr\rangle\langle gg\cdots g0|\\
&+e^{iU_{N-1,N}t}|gg\cdots er\rangle\langle gg\cdots e0|+\cdots\\
&+\left[\exp{i(\sum_{i=1}^{N-2}U_{iN})t}\right]|ee\cdots gr\rangle\langle ee\cdots g0|\\
&+\left[\exp{i(\sum_{i=1}^{N-1}U_{iN})t}\right]|ee\cdots er\rangle\langle ee\cdots e0|)\\
&+\Omega_2(t)e^{-i(\Delta t+\varphi_2)}(|gg\cdots gr\rangle\langle gg\cdots g1|\\
&+e^{iU_{N-1,N}t}|gg\cdots er\rangle\langle gg\cdots e1|+\cdots\\
&+\left[\exp{i(\sum_{i=1}^{N-2}U_{iN})t}\right]|ee\cdots gr\rangle\langle ee\cdots g1|\\
&+\left[\exp{i(\sum_{i=1}^{N-1}U_{iN})t}\right]|ee\cdots er\rangle\langle ee\cdots e1|)+{\rm H.c.}.
\end{aligned}
\end{equation}
Under the conditions $\Delta\approx\sum_{j=2}^{N}U_{1j}\approx\sum_{i=1}^{N-1}U_{iN}$ and $\{U_{1j}, U_{iN}\}\gg\Omega_0(t), \Omega_1(t), \Omega_2(t)$, the Hamiltonian in Eq.~(\ref{HrotN}) can be reduced to an effective four-level configuration as
\begin{equation}\label{HeffN}
\begin{aligned}
H_{\rm eff}(t)&=\Omega_0(t)|ee\cdots er\rangle\langle ge\cdots er|\\
&+\Omega_1(t)e^{-i\varphi_1}|ee\cdots er\rangle\langle ee\cdots e0|\\
&+\Omega_2(t)e^{-i\varphi_2}|ee\cdots er\rangle\langle ee\cdots e1|+{\rm H.c.},
\end{aligned}
\end{equation}
which is necessary in constructing the $N$-qubit controlled gates within our scheme. Using parametric settings similar to those Eqs.~(\ref{H_effdb}) and (\ref{H1_effdb}) and the Morris-Shore transformation, the effective Hamiltonian in Eq.~(\ref{HeffN}) can be recast in the form
\begin{equation}\label{HeffNbd}
\begin{aligned}
H_{\rm eff}(t)&=\Omega(t)e^{i\varphi_2}|b\rangle\langle ee\cdots er|\\
&+\Omega_0(t)|ge\cdots er\rangle\langle ee\cdots er|+{\rm H.c.}
\end{aligned}
\end{equation}
with the bright and dark states
\begin{equation}\label{brightbasisN}
\begin{aligned}
|b\rangle&=\sin{\frac{\theta}{2}}e^{i\varphi}|ee\cdots e0\rangle-\cos{\frac{\theta}{2}}|ee\cdots e1\rangle,\\
|D_1\rangle&=\cos{\frac{\theta}{2}}|ee\cdots e0\rangle+\sin{\frac{\theta}{2}}e^{-i\varphi}|ee\cdots e1\rangle,
\end{aligned}
\end{equation}
respectively, where the time-independent dark state $|D_1\rangle$ constitutes the first dark path. Similar to the preceding few-qubit gates, the second dark path state can be constructed as
\begin{equation}\label{darkpathN}
\begin{aligned}
|D_2(t)\rangle&=\cos{u(t)}\cos{v(t)}e^{i\varphi_2}|b\rangle-i\sin{u(t)}|ee\cdots er\rangle\\
&-\cos{u(t)}\sin{v(t)}|ge\cdots er\rangle,
\end{aligned}
\end{equation}
with the same time-dependent setting about $u(t)$ and $v(t)$ as in the few-qubit controlled gates. The time-dependant parameters $\Omega(t)$ and $\Omega_0(t)$ are determined by Eq.~(\ref{Om_Omtildetwo}) as well.

\begin{figure}[htbp]
\centering
\includegraphics[width=0.85\linewidth]{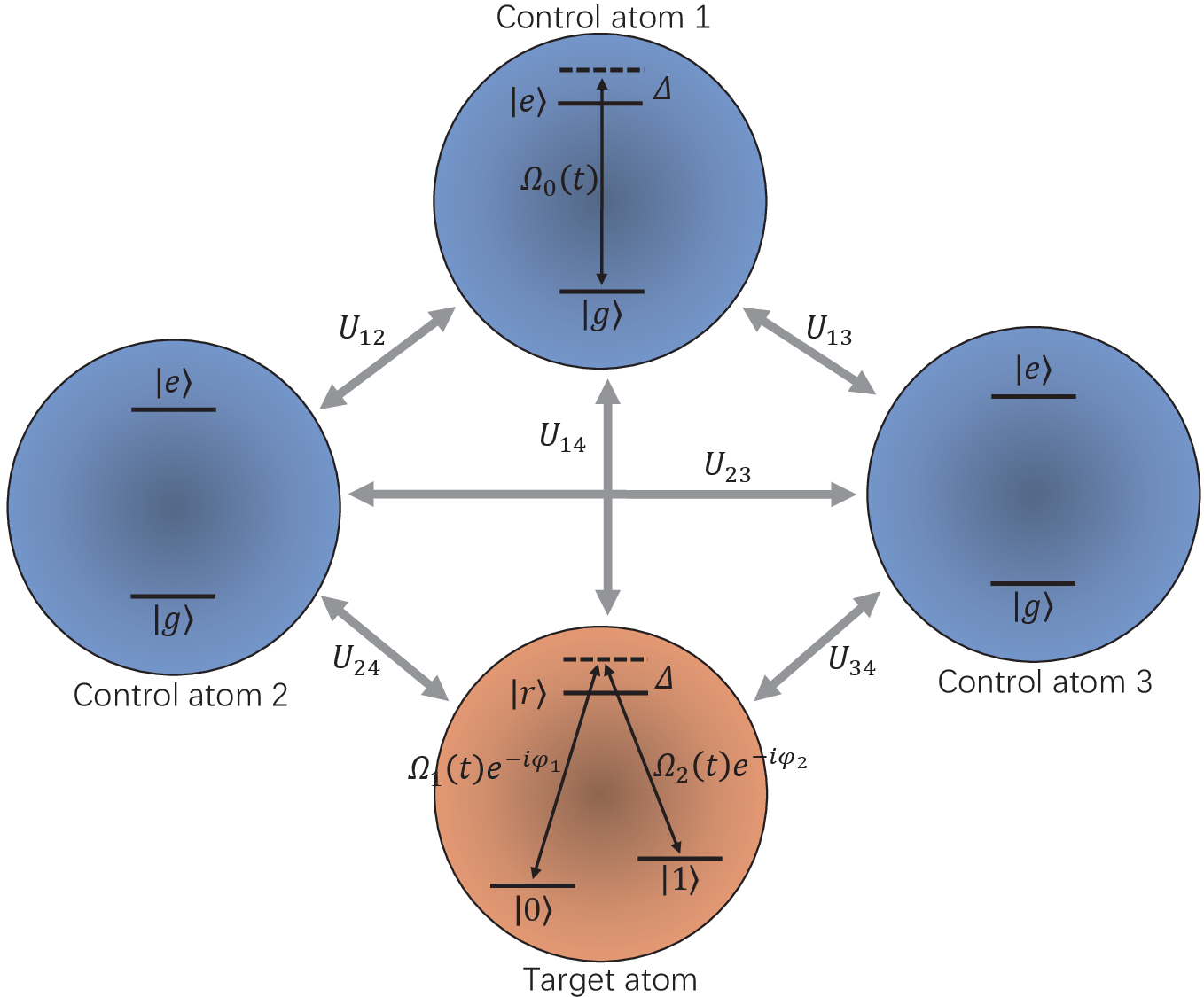}
\caption{Sketch of four coupled high-lying Rydberg atoms, where only the first control atom and the target atom are under driving. Here $U_{i4}$ is the van der Waals interaction between the $i$th control atom and the target atom; $U_{ij}$ ($i<j$) is the interaction between the $i$th and $j$th control atoms.}\label{model4}
\end{figure}

Arbitrary holonomic $N$-qubit controlled gates can be implemented by using the multipulse single-loop technique~\cite{Herterich2016Singleloop}, which results in the holonomic matrix of the same formation in Eq.~(\ref{Utau}). In the gate subspace $\{|ee\cdots e0\rangle, |ee\cdots e1\rangle\}$, the unitary transformation can be written as $U(\theta,\varphi,\gamma)=U(\tau,0)=|ee\cdots e\rangle\langle ee\cdots e|\otimes e^{i(\gamma/2)}e^{-i(\gamma/2)\vec{n}\cdot\vec{\sigma}}$, which indicates that the rotation of the target atom can be launched when $N-1$ control atoms are prepared at the Rydberg state $|e\rangle$. Different multiqubit geometric gates can be realized through modulating the amplitudes and phases of driving fields, such as the $C_N$-NOT gate $U(\pi/2,0,\pi)$ and the $C_N$-Z gate $U(0,0,\pi)$. In sharp contrast to the previous $N$-qubit scheme in a Rydberg-atom system~\cite{Xing2021Realization}, the required holonomic transformation in our scheme is constructed by only one step, which means a great reduction in the parametric manipulation.

We now consider an example with $N=4$ (see Fig.~\ref{model4}) to estimate the feasibility of our theoretical scheme. We use the high-lying Rydberg atoms with $n=103$ that are coupled through van der Waals interactions with the coefficient $C_6\approx2\pi\times1\times10^5$ GHz $\mu m^6$~\cite{Cano2014Multiatom,Sun2021Onestep,Sun2022Onestep}. To realize the four-qubit gates, we can set $d_{12}=d_{13}=d_{24}=d_{34}\approx9 \mu$m and $d_{14}=8 \mu$m, which correspond to the interaction strengths $U_{12}=U_{13}=U_{24}=U_{34}\approx2\pi\times200$ MHz and $U_{14}\approx2\pi\times400$ MHz, respectively. According to the far-off-resonant condition for the effective Hamiltonian in Eq.~(\ref{HeffN}), the detuning and Rabi frequencies can be tuned as $\Delta=2\pi\times800$ MHz and $\{\Omega_0(t),\Omega_1(t),\Omega_2(t)\}\approx2$ MHz, respectively.

\section{Conclusion}\label{Conclusion}

In summary, we have constructed an arbitrary nonadiabatic holonomic $N$-qubit controlled gate based on the dark paths in the Rydberg-atom system. Our scheme relies only on the Rydberg-like interaction $U|er\rangle\langle er|$ with an accessible coupling strength $U$ in experiments, irrespective of its construction or origin. It features with an extraordinary robustness against both external noises and systematic errors by virtue of significant modifications over both conventional nonadiabatic holonomic quantum computation and dark paths. The Rydberg- atom system of arbitrary size $N$ could be effectively described with a four-level configuration as long as the common detuning of driving fields on the target atom and the first control atom $\Delta$ is nearly resonant with both the sum of coupling strengths between the first control atom and the other atoms $U_{1j}$ and the sum of those between the target atom and the other atoms $U_{jN}$. Thereby, the holonomic transformation in our NHGTD scheme for geometric controlled gates of arbitrary type and arbitrary size can be inversely engineered by dark paths and the multipulse single-loop method. It is basically a one-step operation and greatly reduces the cost of parametric modulation in existing schemes. With the high performance in gate fidelity and the convenient scalability to more qubits, our scheme is of interest in the pursuit of high-speed and large-scale quantum computation.

\section*{Acknowledgments}

We acknowledge financial support from the National Natural Science Foundation of China (Grant No. 11974311).

\bibliographystyle{apsrevlong}
\bibliography{ref1}

\end{document}